\documentclass[9pt,sigconf]{acmart}

\usepackage{cleveref}
\crefname{equation}{Eq.}{Eqs.}
\crefname{figure}{Fig.}{Figs.}
\crefname{section}{Sec.}{Secs.}
\crefname{chapter}{Ch.}{Chs.}
\crefname{appendix}{Appx.}{Appxs.}
\newcommand{\ttt}{\texttt}

\setcopyright{none}

\renewcommand\footnotetextcopyrightpermission[1]{}
\begin{document}
\title[Automatically Tuning GCC to Optimize Applications on Embedded Systems]{Automatically Tuning the GCC Compiler to Optimize the Performance of Applications Running on Embedded Systems}

\author{Craig Blackmore, Oliver Ray, Kerstin Eder}
\affiliation{University of Bristol, Merchant Venturers Building, Woodland Road\\Bristol, BS8 1UB, United Kingdom}
\email{{firstname.lastname}@bristol.ac.uk}

\renewcommand{\shortauthors}{}

\begin{abstract}
This paper introduces a novel method for automatically tuning the selection of compiler flags to optimize the performance of software intended to run on embedded hardware platforms. We begin by developing our approach on code compiled by the GNU C Compiler (GCC) for the ARM Cortex-M3 (CM3) processor; and we show how our method outperforms the industry standard \ttt{-O3} optimization level across a diverse embedded benchmark suite. First we quantify the potential gains by using existing iterative compilation approaches that time-intensively search for optimal configurations for each benchmark. Then we adapt iterative compilation to output a single configuration that optimizes performance across the entire benchmark suite. Although this is a time-consuming process, our approach constructs an optimized variation of \ttt{-O3}, which we call \ttt{-Ocm3}, that realizes nearly two thirds of known available gains on the CM3 and significantly outperforms a more complex state-of-the-art predictive method in cross-validation experiments. Finally, we demonstrate our method on additional platforms by constructing two more optimization levels that find even more significant speed-ups on the ARM Cortex-A8 and 8-bit AVR processors.
\end{abstract}



\maketitle
\section{Introduction}
Modern compilers offer a range of optimization levels that are intended to progressively improve the execution time of programs at the expense of increased compile time, code size and/or conformance to software standards. The most famous is the \ttt{-O3} optimization level, provided by the GCC compiler~\cite{gcc} (and its more recent competitor Clang~\cite{clang}), which is widely used as the optimization of choice in industry.
Previous work~\cite{fursin2011} has shown that \ttt{-O3} is far from optimal in many cases. By selectively enabling or disabling compiler flags that control optimization settings, the compiler can be fine-tuned to improve the performance of a given program and target platform. This leads to significant gains without the need to modify the underlying source code nor the compiler itself.

Finding effective \emph{configurations} of compiler flags is, however, a hard task due to the large number of flags available and complex and often unknown interactions between them. An exhaustive search would take infeasibly long. Furthermore, the optimal configuration is dependent on the target program and platform.

Existing work uses random sampling or more complex \emph{iterative compilation}~\cite{pan2006} methods (which evaluate the performance of a given program compiled with a large number of different configurations) to search for configurations that improve the performance of a target program. This is a time consuming task which must be repeated for each program and platform pair. The slow search time motivated other studies to use iterative compilation to train machine learning approaches to predict good configurations more quickly at the cost of reduced accuracy for an unseen program.

This paper shows how iterative compilation methods can be adapted to discover a single configuration tailored to a target platform in order to outperform the default optimization levels provided by the compiler. In contrast to previous methods, which perform a new search for each program, we search for a single configuration that enhances the overall performance of a wide range of benchmarks on a given platform. This single configuration can then simply be used in place of \ttt{-O3} with no further effort from the compiler writer or application developer.

We develop our approach on the industry standard GCC compiler and the STM32VLDISCOVERY embedded system development board which features an ARM Cortex-M3 (CM3) 32-bit processor that is a popular choice of processor for Internet of Things platforms~\cite{mbed}.
We use the state-of-the-art open source Bristol/Embecosm Embedded Benchmark Suite (BEEBS)~\cite{beebs2013} to measure the effects of different configurations on a diverse set of programs.

First we perform an investigatory study to quantify the potential gains, by using state-of-the-art iterative compilation methods that time-intensively search for optimal configurations for each benchmark. We then propose a practical method to automatically construct a single configuration, \ttt{-Ocm3}, that gives near-optimal speed-up across the benchmarks. In addition, we analyze the effects of two of the flags which our method determines should be removed from \ttt{-O3} on our target architecture and explain in detail why disabling them does indeed improve performance.

We evaluate \ttt{-Ocm3} further by using 10-fold cross-validation to show that our approach generalizes well to previously unseen test cases and outperforms a more complex state-of-the-art machine learning approach~\cite{fursin2011}. Our results suggest that it is best to use \ttt{-Ocm3} in place of \ttt{-O3} on the CM3 in order to maximize performance.

Finally, we demonstrate the benefit of our method on two additional embedded platforms. The first platform is the BeagleBone development board which has an ARM Cortex-A8 (CA8) 32-bit processor that has featured in many mobile devices and is much more complex than the CM3. The second platform is the ATmega328P microcontroller which features an AVR 8-bit processor that is commonly used in Arduino devices and is much simpler than the CM3. We construct two new optimization levels, \ttt{-Oca8} and \ttt{-Oavr}, that outperform \ttt{-O3} on the CA8 and AVR respectively.

The rest of the paper is structured as follows. In~\cref{sec:bg} we give the technical background that is relevant to our work. Then we present our investigatory study~(\cref{sec:invest}). This is followed by the development of our approach to construct a new optimization level~(\cref{sec:results}) and our cross-validation experiments~(\cref{sec:valid}). Then we test the approach on two additional platforms~(\cref{sec:plat}) and we discuss the wider context of this research by summarizing related work~(\cref{sec:rel}). Finally, we discuss conclusions and future work~(\cref{sec:conclusion}).

\section{Background}
\label{sec:bg}

\sloppy
To make this paper self-contained, this section gives a brief outline of the standard optimizations available in GCC (\cref{sec:bg:gcc}) followed by a summary of compiler tuning techniques for identifying effective compiler settings (\cref{sec:bg:ic,sec:bg:ml}) and a brief introduction to the BEEBS benchmark suite (\cref{sec:bg:beebs}).

\subsection{Standard Optimization Levels}
\label{sec:bg:gcc}

Modern compilers provide standard optimization levels which enable a predefined set of optimizations. GCC provides \ttt{-O0}, \ttt{-O1}, \ttt{-O2} and \ttt{-O3} which enable an increasing set of optimizations at the expense of code size~\cite{gcc}. There is also \ttt{-Os} which is similar to \ttt{-O2} except it disables optimizations expected to increase code size. Finally, \ttt{-Ofast} applies additional optimizations to \ttt{-O3} that do not conform to industry standards (e.g. IEEE floating point) and therefore compromise precision, compatibility and reproducibility. Although these optimization levels are convenient for the user, better settings can often be found with extra effort (\cref{sec:bg:ic} below).

\subsection{Iterative Compilation}
\label{sec:bg:ic}
Iterative compilation~\cite{kisuki1999} methods compile a target program with several different compiler configurations and evaluate the performance of each resulting compilation in order to find a good one. This is a time-consuming task that must be repeated for each new program and platform combination but in practice it yields significant gains. There are several approaches for selecting which configurations to test in iterative compilation. Two of the most popular methods are Random Iterative Compilation (RIC)~\cite{fursin2011} and Combined Elimination (CE)~\cite{pan2006}. 

\textbf{Random Iterative Compilation (RIC)} uses straight-forward random sampling of compiler flags to construct a set of configurations for evaluation.

\textbf{Combined Elimination (CE)} seeks to analyze the effect of each flag relative to an initial baseline, which has all flags enabled, and continually updates the baseline by disabling the flag that has the largest negative impact on performance. We briefly give the CE algorithm described in~\cite{pan2006}. The algorithm uses the Relative Improvement Percentage (RIP) to measure the impact of a given flag in relation to a given configuration and target program. Let $F_1,F_2,...,F_n$ be the set of available compiler flags. The impact of flag $F_i$ relative to the baseline configuration $B$ is calculated by the following:
$$RIP_{B(F_i)} = \frac{T_{B(F_i = 0)} - T_B}{T_B} * 100$$
where $T_B$ is the execution time of the target program when compiled with configuration $B$ and $T_{B(F_i = 0)}$ is the execution time given by the same configuration with flag $F_i$ disabled. The algorithm proceeds as follows:

\begin{enumerate}
\item Let $S=\{F_1,F_2,...,F_n\}$ be the optimization search space and $B=\{F_1=1,F_2=1,...,F_n=1\}$ be the baseline configuration with all flags enabled.
\item Calculate the $RIP_{B(F_i)}$ for each flag $F_i \in S$.
\item Let $X=\{X_1,X_2,...,X_m\}$ be the set of flags with negative RIPs sorted in ascending order such that $X_1$ has the most negative RIP.
\item If $X=\emptyset$ then terminate with $B$ as the final configuration.
\item Remove $X_1$ from $S$ and $X$ and let $B=B(X_i=0)$.
\item For $i$ = 2 to $m$ recalculate $RIP_{B(X_i)}$ and if $RIP_{B(X_i)} < 0$ remove $X_i$ from $S$ and $X$ and let $B=B(X_i=0)$.
\item Goto step 2.
\end{enumerate}

Pan et al. \cite{pan2006} showed that CE outperforms other iterative compilation approaches such as Optimization-Space Exploration (OSE)~\cite{triantafyllis2003} and Statistical Selection (SS)~\cite{pinker2004}. Although Cavazos et al. \cite{cavazos2007} later concluded that RIC outperforms CE on an AMD Athlon case study, this does not appear to hold on our embedded system study~(\cref{sec:invest:ceVSrandom} later on).

\subsection{Machine Learning Approaches}
\label{sec:bg:ml}

Due to its time-intensive nature, it is clearly infeasible to use full iterative compilation every time a programmer wants to compile a new program. This motivated other studies to use iterative compilation data to train machine learning based approaches that seek to predict a suitable configuration to optimize a given target program. Typically, these methods train a model which takes an input that describes characteristics of the target program and outputs a predicted configuration.
These methods exhibit a trade-off between the time taken to find a solution and the quality of that solution.

Many predictive compiler tuning approaches rely on feature vectors of statistical aggregates that summarize characteristics of the target program code. These methods seek to correlate program features with effective configurations but finding the most relevant features is non-trivial.

Milepost~\cite{fursin2011} used 1-nearest-neighbor (1NN) and decision tree approaches to train and test models based on RIC data and a feature vector of 56 features. The study focused on optimizing the most time consuming function of each program and concluded that their 1NN probabilistic approach performed best. Given a target program, this approach identifies the training program with the closest feature vector based on the most time consuming function and uses its RIC results to predict a configuration for the target program.

Kulkarni et al. \cite{kulkarni2012} used a technique called Neuro-Evolution for Augmenting Topologies (NEAT)~\cite{stanley2002} to train a neural network to predict performance enhancing optimization sequences for the Jikes RVM~\cite{jikes2005} Java compiler. This approach generates an initial population of neural networks and uses a genetic algorithm to evaluate and evolve new neural networks. 
Sher et al. \cite{sher2014} also used NEAT to learn neural networks that predict optimization sequences for LLVM.

More recently, Blackmore et al. \cite{Blackmore2015} proposed a logic based machine learning approach that seeks to automatically discover relevant features for predicting effective compiler flags.

\subsection{BEEBS}
\label{sec:bg:beebs}

This study uses the 84 benchmarks of the Bristol/Embecosm Embedded Benchmark Suite (BEEBS)~\cite{beebs2013}, which to our knowledge is the largest collection of free open source benchmarks available for embedded systems.\footnote{Six programs are no longer in the master branch as their license status could not be confirmed and BEEBS requires all benchmarks to be under GPL.}
The benchmarks cover a wide range of characteristics as demonstrated in~\cite{Blackmore2015} and were produced in response to the lack of freely available benchmarks for resource limited bare-metal embedded systems such as the CM3 and AVR.\footnote{Although the CA8 is able to run Linux, we run all of the benchmarks on bare-metal to prevent the OS from interfering with timings.} Other existing benchmark suites have fewer benchmarks and are unsuitable for this study due to their reliance on an OS and/or file system being present. Some of the BEEBS programs were in fact derived and adapted from the MiBench~\cite{mibench}, WCET~\cite{wcet} and DSPstone~\cite{dspstone} suites.

Each BEEBS benchmark consists of at least one source file containing the benchmark itself plus another file \ttt{main.c} which controls the number of times the benchmark is run according to a repeat factor. The repeat factor is used to produce a runtime long enough to obtain reliable measurements and it also enables BEEBS to target a wide range of platforms which may execute particular benchmarks considerably faster or slower than other systems. Programs that run too fast may need to be looped tens of thousands of times in order to produce a long enough runtime. In these cases, the loop overhead may account for most of the measurement.

Most of the benchmarks require test input data on which to operate. In reality, the input data would not be known at compile-time and would typically be supplied via command-line parameters, data files or an input stream from a device (e.g. sensor). To make a fair comparison between different compilations of the same benchmark, the input data must be fixed. However, BEEBS targets bare-metal embedded systems which have no command line or file handling support for providing input files or parameters, therefore the input is fixed by hard-coding it into the source code.

Note that the latest version of BEEBS includes some technical improvements that were made (as part of the investigatory study described in the next section) in order to prevent the compiler from over-optimizing a benchmark based on its advance knowledge of the input data on which that benchmark will operate.

\section{Quantifying Potential Gains}
\label{sec:invest}

\begin{figure*}[ht!]
\begin{center}
\includegraphics[width=\textwidth,keepaspectratio]{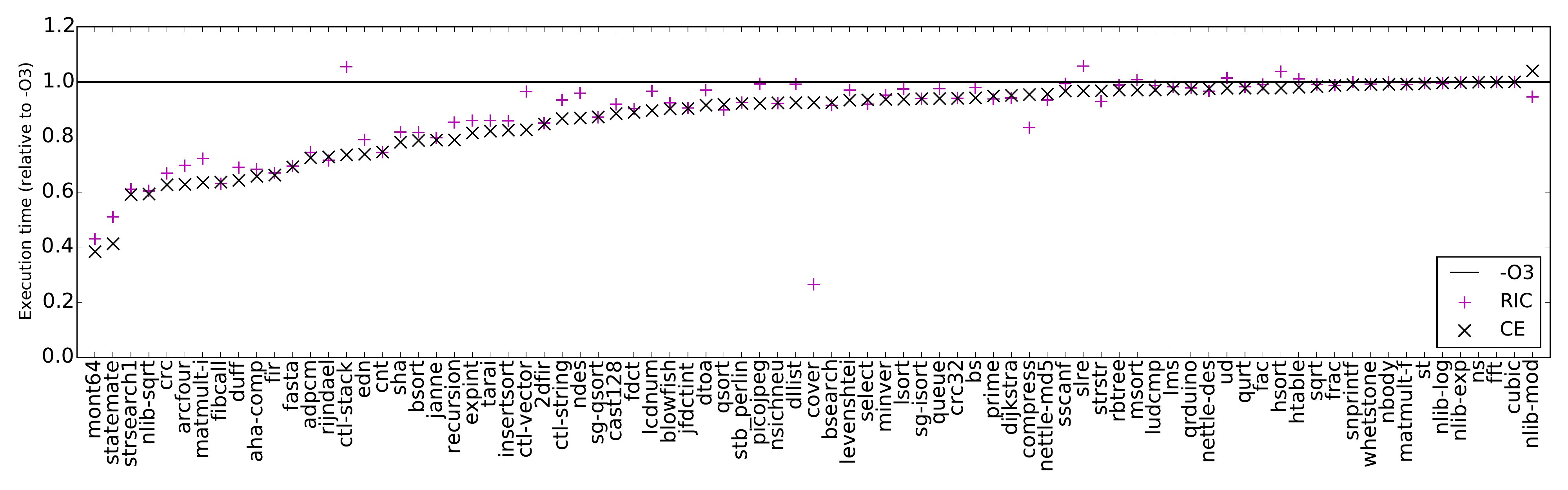}
\end{center}
\caption{Best execution time achieved by RIC (1000 configurations) and CE}
\label{fig:ceResultsPerBmark}
\end{figure*}

The aim of this section is to find the potential gains available for each benchmark in BEEBS by using RIC and CE to search as exhaustively as possible for optimal configurations.\footnote{This analysis excludes three programs that do no run on the STM32VLDISCOVERY.}

First, we identify and fix an oversight in the design of BEEBS in order to increase confidence in our results~(\cref{sec:invest:beebs}). We then intensively search for potential gains using RIC and CE (\cref{sec:invest:data}), contrast the performance of these two methods~(\cref{sec:invest:ceVSrandom}) and summarize the potential gains (\cref{sec:invest:gains}).

\subsection{BEEBS Data Initialization}
\label{sec:invest:beebs}

In completing this study, we identified and fixed an oversight in the design of BEEBS in order to increase the reliability of our experiments and future work. We found that the compiler was able to `over-optimize' given knowledge of test input data necessarily hard coded into benchmarks. This is a conceptual flaw that might also affect other benchmark suites.

We have edited BEEBS to eliminate cases where it was possible for the compiler to optimize based on input data.\footnote{Our changes are now in the master branch of BEEBS (http://beebs.eu)} This was done by using an \ttt{initialise\_benchmark} function which initializes any input data required by the benchmark. The \ttt{initialise\_benchmark} function is defined outside of \ttt{main.c}, but is called from within \ttt{main.c}. As long as link-time optimization is disabled, the knowledge that \ttt{initialise\_benchmark} is called in \ttt{main.c} cannot be used in the optimization of the other source files. Benchmarks for which input data was given by global variables or arrays did not need adjusting because the compiler cannot assume that the globals are not changed elsewhere in the program.

Over-optimization led to 1\% of overall gains seen in our preliminary experiments. Ten of the benchmarks gained over 5\% advantage from having input data exposed to the compiler.

For example, the compiler was able to over-optimize \ttt{expint} (which calculates exponential integrals) based on constant input data to a key function in the benchmark. With inputs as constants, we found a configuration that reduced the execution of \ttt{expint} by 18\% compared to \ttt{-O3}, but without these inputs available for optimization the reduction was a smaller 8\%.

The rest of this study proceeds using our improved version of BEEBS.

\subsection{Generating Data}
\label{sec:invest:data}
To search for potential gains on our target architecture we focused on 133 flags available when compiling for the CM3 in GCC 4.9.3. This includes 26 flags not enabled by \ttt{-O3} and excludes flags that do not follow the standards, produce incompatible code, reduce precision, require additional profiling information or are purely intended for C++ or debugging.

\label{sec:invest:random}
For our RIC data we used a similar method to~\cite{fursin2011} and~\cite{Blackmore2015}. We generated a random sample of 1000 configurations by selecting \ttt{-O1}, \ttt{-O2} or \ttt{-O3} with probability $p(\frac{1}{3})$ and enabling each flag with probability $p(\frac{1}{2})$. Our CE data was generated using the original CE algorithm~\cite{pan2006} as described in~\cref{sec:bg:ic}.

To improve the efficiency of each method, we store the md5 hash of each compiled binary along with its performance measurement. If any future compilation has the same hash, we use the previously cached performance rather than re-executing the binary.

\subsection{Performance of RIC vs CE}
\label{sec:invest:ceVSrandom}

This section compares the performance of RIC and CE in terms of best configuration found per benchmark and time taken to find good configurations. We show that overall, CE outperformed RIC but we also give insights as to why RIC occasionally finds better configurations.

The best execution times achieved by RIC and CE are shown in \cref{fig:ceResultsPerBmark}. On average the two methods performed 11\% and 13\% better than \ttt{-O3} respectively. Combined Elimination outperformed RIC on three quarters of the benchmarks. There were six benchmarks for which RIC was unable to outperform \ttt{-O3} despite CE finding better configurations. In particular, the execution time of ctl-stack was improved by over 25\% by CE while RIC performed comparable to \ttt{-O3}.

Conversely, RIC performed significantly better than CE on three benchmarks -- \ttt{cover}, \ttt{compress} and \ttt{newlib-mod}. Analysis of the RIC results for \ttt{cover} showed that two flags (\ttt{-fivopts} and \ttt{-ftree-ch}) were always disabled in the best configurations. Further experiments showed that exclusively disabling one of these flags degraded performance and it was in fact the combination of both flags being disabled that led to improved performance. This is a dependency between the two flags which the CE algorithm is unable to capture due to the way it considers a single flag at a time.

While CE does not completely disregard dependencies (each decision to enable or disable a flag is dependent on the current baseline configuration) it only considers the effect of a single flag at a time, rather than toggling multiple flags at once. Allowing all single and pairs of flags to be toggled increases the search space exponentially but as a compromise, the CE algorithm could be modified to consider groups of flags with known dependencies (although finding these dependencies is non-trivial~\cite{pallister2013}). Further work is required to determine whether \ttt{compress} and \ttt{newlib-mod} also exhibit dependencies between flags that CE was unable to capture.

\begin{figure}[hb!]
\begin{center}
\includegraphics[width=\columnwidth,keepaspectratio]{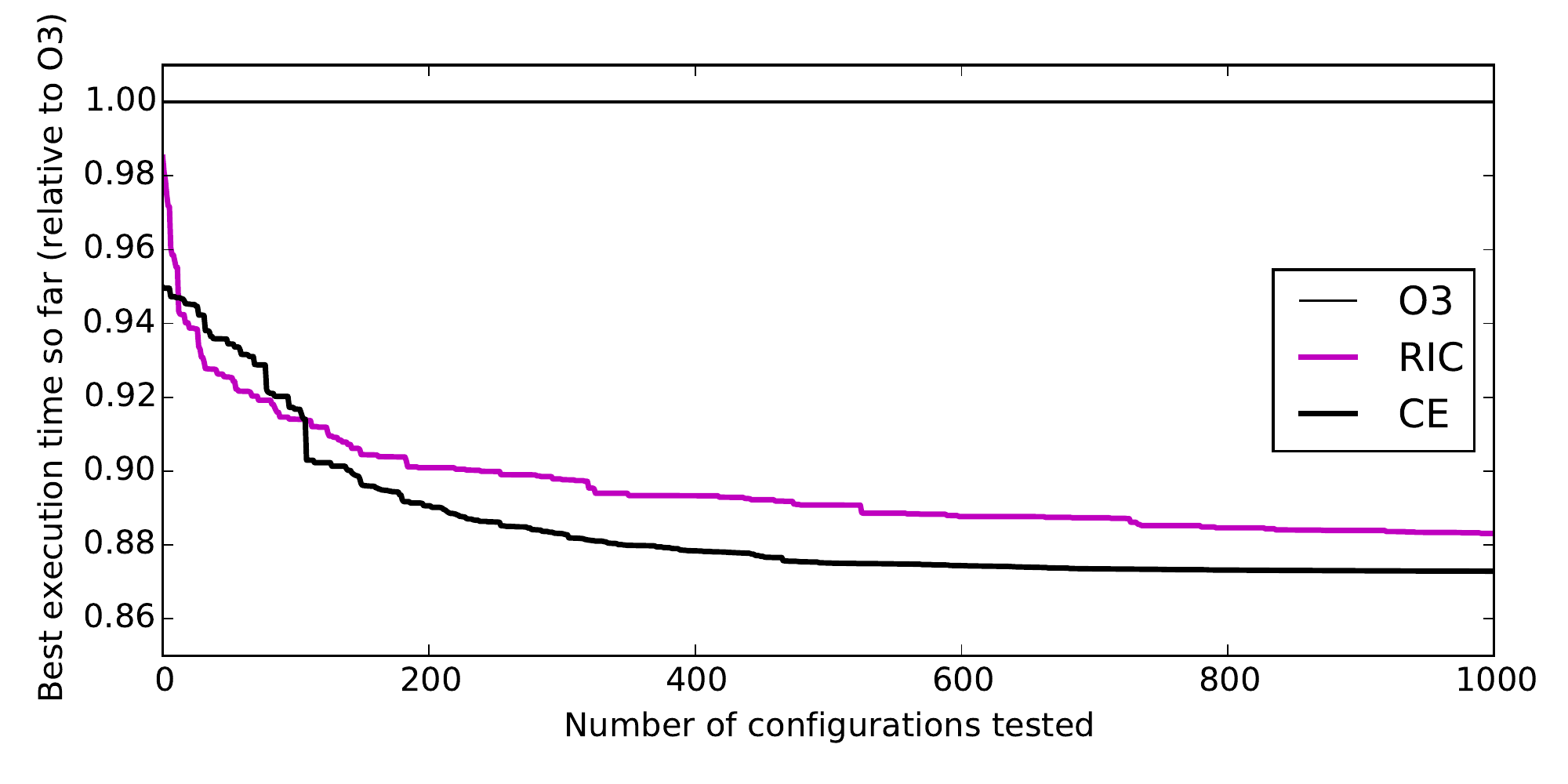}
\end{center}
\caption{Average of best execution time for each benchmark after each configuration tested by RIC (1000 configurations) and CE}
\label{fig:ceTimeSeries}
\end{figure}

The real value of CE becomes apparent when analyzing the amount of time each method takes to find good configurations. This is shown by plotting the average of the current best performance achieved on each benchmark after each configuration is tested~(\cref{fig:ceTimeSeries}). In calculating the average, the performance of each benchmark is floored with \ttt{-O3} in order to compare with previous work~(\cref{sec:rel:iter}).

Combined Elimination overtakes RIC after 108 configurations have been tested and stays in the lead for the remaining iterations. Note that CE takes 134 configurations to test the initial baseline and each of the 133 flags. At configuration 108, a single flag (\ttt{-ftree-loop-if-convert}) is disabled, which has a strong impact on performance. This flag will be analyzed further in~\cref{sec:tree}.

Our RIC experiments were terminated after 1000 configurations due to time restraints, but the trajectory suggests it would take much longer for RIC to match the performance achieved by CE. Overall, RIC iterative compilation took 7.5 days to run and CE took 2.5 days.

\subsection{Summary of Potential Gains}
\label{sec:invest:gains}

\begin{figure*}[ht!]
\begin{center}
\includegraphics[width=\textwidth,keepaspectratio]{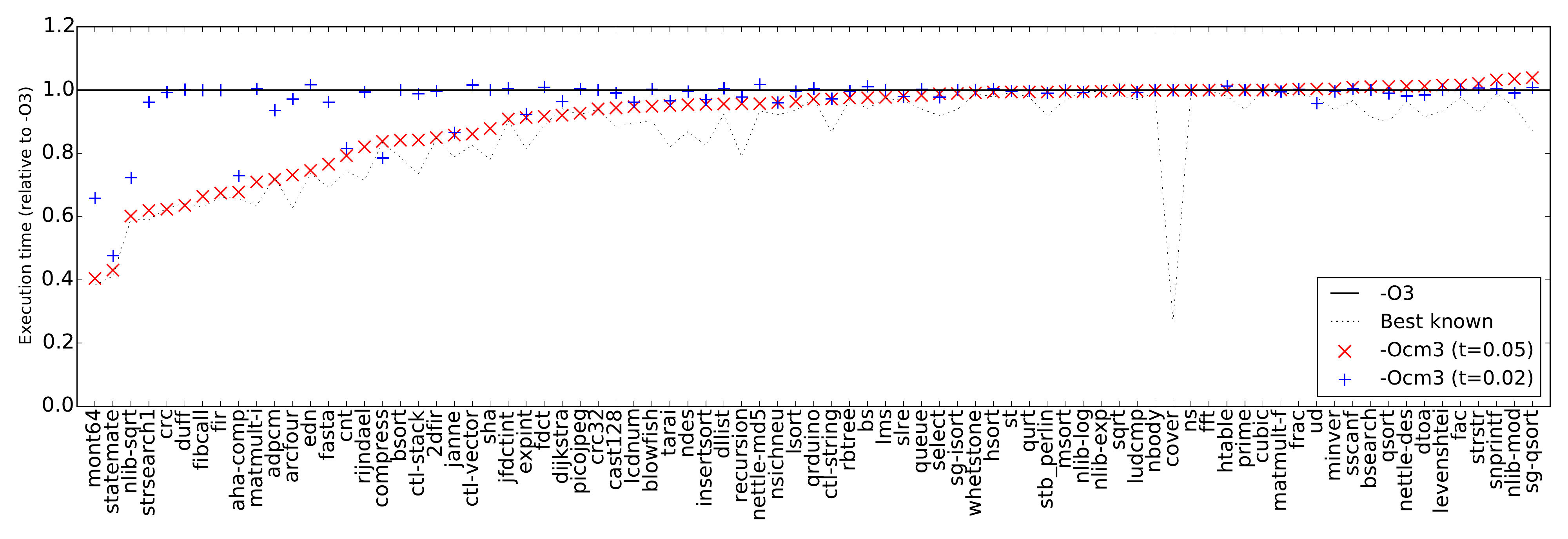}
\end{center}
\caption{Trade-off between thresholds for constructing \ttt{-Ocm3}}
\label{fig:ocm3thresh}
\end{figure*}

To quantify the potential gains available on the CM3 we take the best known configuration found by either RIC or CE for each benchmark~(\cref{fig:ceResultsPerBmark}). This gives an overall improvement of 14\% compared to \ttt{-O3}. The best known configuration for each program provides a target with which to compare any proposed method, such as the one we introduce in the next section, that aims to improve upon \ttt{-O3}.

\section{Constructing a New Optimization Level}
\label{sec:results}

In this section we propose a general methodology for adapting existing iterative compilation methods in order to find a single configuration to optimize the performance of a whole benchmark suite~(\cref{sec:method:ceWhole}). We demonstrate the method in practice by applying it to CE to construct \ttt{-Ocm3} based on 81 programs from BEEBS on the CM3. We discuss the effect of our method's threshold parameter which controls the trade-off between performance gains on some programs in exchange for small losses on others~(\cref{sec:results:ocm3}). Finally, we analyze two flags which our method suggests should always be disabled for the CM3 to determine why removing them enhances performance~(\cref{fig:results:badFlags}).

\subsection{Adapting Iterative Compilation}
\label{sec:method:ceWhole}

Existing iterative compilation techniques search for a single configuration to improve the performance of a particular program. By changing the goal from enhancing a target program's performance to maximizing the overall performance of a suite of benchmarks we can adapt iterative compilation methods to find a single configuration tailored to the target platform. The aim of our approach is to find a single configuration that improves overall performance without having a significant negative impact on any one program.

We demonstrate our new strategy by building the creation of optimization levels into the CE algorithm such that the final configuration is the new optimization level itself.

Intuitively, the method starts with \ttt{-O3} as its baseline configuration and continually enables or disables the next flag which gives the biggest improvement across \emph{all} benchmarks while not causing any one benchmark to perform worse than a threshold $t\%$ of \ttt{-O3}. The result is a configuration that performs at least within $t\%$ of \ttt{-O3} or better for each benchmark. Threshold $t\%$ controls the trade-off between performance gains and loses which will be explored further in \cref{sec:results:ocm3}.

We test our approach on CE by making the following changes to the original algorithm~(\cref{sec:bg:ic}):

\begin{itemize}
\item Instead of targeting the performance of a single program, we target the overall performance of the benchmark suite running on a given platform.
\item Rather than starting with a baseline configuration of all flags enabled and then selectively disabling flags, our baseline configuration is \ttt{-O3} and we can either disable flags that are in \ttt{-O3} or enable flags that are not in \ttt{-O3}. Any baseline configuration upon which the user wishes to improve can be chosen here.
\item{As soon as a configuration causes a program to perform $t\%$ worse than \ttt{-O3}, the remaining tests for that configuration are skipped as it will not satisfy the requirement that performance must be at least within $t\%$ of \ttt{-O3} or better. This increases the efficiency of the search by avoiding unnecessary evaluations.}
\item{As in~\cref{sec:invest:data}, to further aid efficiency of the search, the md5 hash of each compiled binary is stored along with its performance measurement. The cached performance is used for any subsequent binary with a matching hash rather than rerunning the program.}
\end{itemize}

In \cref{sec:results:ocm3} we analyze the results of applying our method to find a single configuration that outperforms \ttt{-O3} on the CM3 and we also highlight how our changes improve the efficiency of the search.

\subsection{Threshold Trade-off in Constructing \ttt{-O\lowercase{cm3}}}
\label{sec:results:ocm3}

Our proposed method for constructing \ttt{-Ocm3} (\cref{sec:method:ceWhole}) was tested with several thresholds from $t=0\%$ to $t=6\%$ using fixed increments of 1\%. The configuration generated by $t=5\%$ gave the best average performance which was 9\% better than \ttt{-O3}. Under this configuration, many of the benchmarks perform close to their best known configuration and only a few perform worse than \ttt{-O3}~(\cref{fig:ocm3thresh}). The worst performing program ran 4\% slower than \ttt{-O3}. Several benchmarks performed as well as the best known configuration.

A more conservative threshold $t=2\%$ performs 3\% better than \ttt{-O3} overall and still manages several improvements with fewer programs performing worse than \ttt{-O3}. There is, then, a trade-off between optimizing some benchmarks in exchange for losses on others.

Under $t=5\%$ our method identifies 20 flags which should be disabled from \ttt{-O3} and three additional flags which should be enabled. We will discuss two of these flags in detail in \cref{fig:results:badFlags}. Our results demonstrate that the gains of \ttt{-Ocm3} created by $t=5\%$ outweigh the losses and would recommend its use on the CM3 instead of \ttt{-O3}. The complete configuration is given in~\cref{fig:ocm3config} (with the flags shown in the order that they were disabled or enabled by our method).

\begin{figure}
\begin{verbatim}
-O3 -fno-tree-loop-if-convert -fno-common
-fipa-pta -fno-sched-interblock -fno-tree-copyrename
-fno-peephole2 -fno-expensive-optimizations
-fno-ipa-sra -fgcse-las -fno-schedule-insns
-fno-tree-loop-distribute-patterns -fno-caller-saves
-fno-optimize-strlen -fno-inline-functions-called-once
-fno-tree-slsr -fno-tree-scev-cprop -funroll-all-loops
-fno-sched-dep-count-heuristic -fno-tree-ccp
-fno-predictive-commoning -fno-ipa-pure-const
-fno-merge-constants -fno-tree-pta
\end{verbatim}
\caption{The \ttt{-O\lowercase{cm}3} Configuration}
\label{fig:ocm3config}
\end{figure}

Our method took 19 hours to run with $t=5$, which is over twice as fast as CE and seven times faster than RIC used in our investigatory study~(\cref{sec:invest}).

\subsection{Analysis of Two Excluded Flags}
\label{fig:results:badFlags}

To explain why some of the flags included in \ttt{-O3} appear to actually reduce performance on the CM3 architecture, we analyze two such flags (\ttt{-fcommon} and \ttt{-ftree-loop-if-convert}) which our method indicates should always be disabled. Although both of these flags are in fact enabled at all optimization levels from \ttt{-O0} upwards, disabling them actually reduces the overall average execution time of BEEBS by 3\% and significantly improves the performance of 13 benchmarks while leaving all the others virtually unaffected.

\subsubsection{-fcommon}
\label{sec:com}
The \ttt{-fcommon} flag controls the placement of uninitialized global variables within object code. As stated in the GCC manual, the flag is provided for compatibility but may lead to a speed or code size penalty on some platforms~\cite{gcc}. Disabling the flag on the Cortex-M3 improves overall execution time by 1\% and has a significant impact on \ttt{statemate} and \ttt{compress} which are improved by 43\% and 16\% respectively. 

The use of \ttt{-fcommon} prevents the compiler from using knowledge that two global variables will share contiguous memory. Such knowledge could be used on the CM3 to exploit Load Multiple Increment After (LDMIA) or Store Multiple Increment After (STMIA) instructions which allow two variables to be loaded or stored in a single instruction.

In more detail, \ttt{-fcommon} allows duplicate definitions of uninitialized global variables across different source files. Each definition of a global variable (including duplicates) appears in the common section of the object code and the linker then chooses which of these definitions to use. Unfortunately, this prevents the compiler from knowing the relative location of global variables, and it cannot optimize based on the assumption that they will occupy contiguous memory.

In contrast, when \ttt{-fcommon} is disabled, each global variable can only be defined once and any other declarations must be qualified with the \ttt{extern} keyword. Each global variable is defined once in the data section of the object code and its location relative to other variables is preserved.

Let us briefly analyze the effect of \ttt{-fcommon} on the following example code:
\begin{verbatim}
    int x,y,z;
    void g() {
      z = x - y;
      x = z * y;
      y = z * x;
    }
\end{verbatim}

Compiling with \ttt{-fcommon} produces twice as many instructions than when it is disabled. Only the disabled version reduces the number of memory instructions by taking advantage of LDMIA and STMIA. In addition, the enabled version uses more than 4 registers which causes a further inefficiency on the CM3 as additional stack operations are required to ensure the extra registers are restored to their original values before the function returns~\cite{armCalling}.

\subsubsection{-ftree-loop-if-convert}
\label{sec:tree}
\begin{figure*}[ht!]
\begin{center}
\includegraphics[width=\textwidth,keepaspectratio]{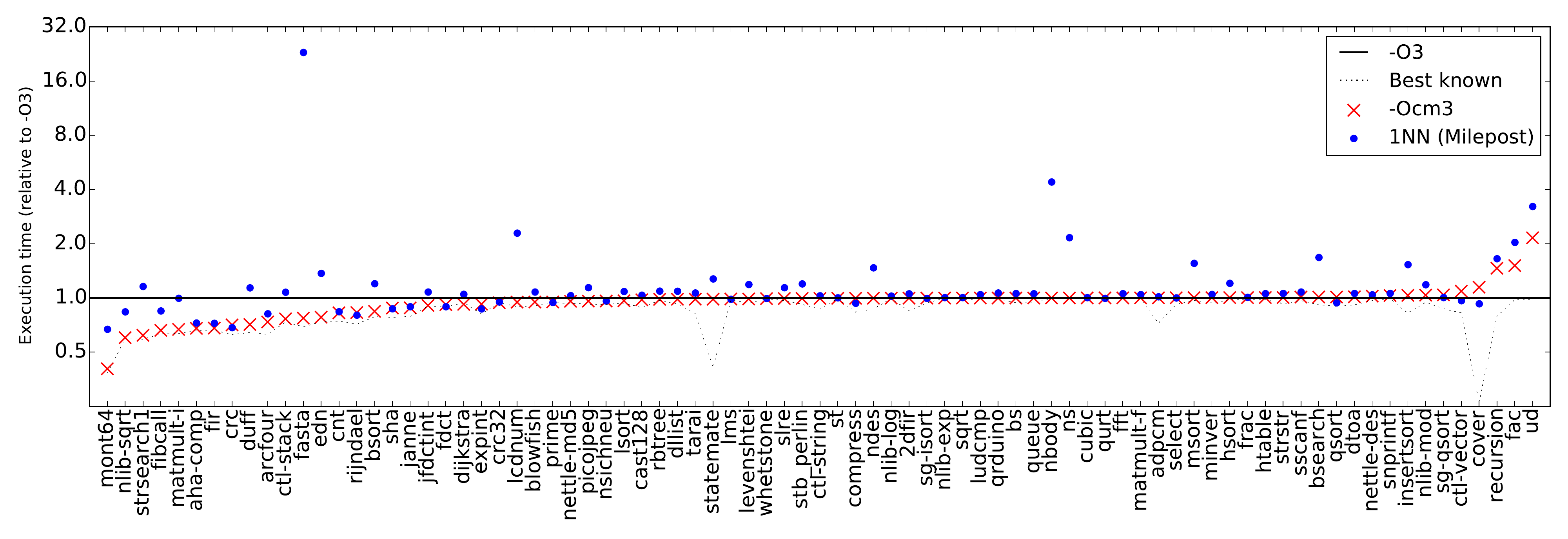}
\end{center}
\caption{10-fold cross-validation of \ttt{-Ocm3} and Milepost 1NN (logarithmic scale)}
\label{fig:test}
\end{figure*}
This flag converts conditional jumps in innermost loops to branchless equivalents in order to improve later vectorization optimizations switched on at \ttt{-O3}~\cite{gcc}. There is no indication in the manual, however, that this flag might degrade performance on a processor such as the CM3 that does not support vectorization. We investigate the impact of this flag further and explain why it does indeed increase runtime on the CM3.

Disabling \ttt{-ftree-loop-if-convert} improves overall execution time by 2\% and significantly improves \ttt{aha-mont} by 50\% and \ttt{newlib-sqrt} and \ttt{aha-compress} by 25\% while not degrading the performance of any remaining benchmark.

Intuitively, this flag removes an if statement and replaces it with code that always executes both the if-true and if-false body and then uses predicated instructions to determine which result(s) should be used. Consider the following if statement found in \ttt{newlib-sqrt}:

\begin{verbatim}
    if(t<=ix) {
        s    = t+r;
        ix  -= t;
        q   += r;
    }
\end{verbatim}

\noindent When \ttt{-ftree-loop-if-convert} is enabled, the code is converted to the following:

\begin{verbatim}
    s2    = t+r;
    ix2  -= t;
    q2   += r;
    ix    = (t<=ix) ? ix2 : ix;
    s     = (t<=ix) ?  s2 : s;
    q     = (t<=ix) ?  q2 : q;
\end{verbatim}

The code produced by \ttt{-ftree-loop-if-convert} always executes the if-true body, but then must execute three more statements to decide which value to use for each variable. In contrast, the original version of the code only executes the if-true body when the condition is true. We anticipate that the second version would perform increasingly well as the proportion of times the if condition evaluates to false increases. Under the default input data for \ttt{newlib-sqrt} the true:false ratio is 1:2.

\subsubsection{Lessons Learned}

This section analyzed two flags in detail to determine why disabling them is beneficial to performance on the CM3. While such manual analysis provides interesting insights it is a time-consuming task that is infeasible to repeat for the very many flags and platforms available. Our new iterative compilation based approach enables the automatic discovery of such important flags for new architectures without the need for in-depth manual analysis.

\section{Cross-validation of \ttt{-O\lowercase{cm}3}}
\label{sec:valid}

In~\cref{sec:results} we used the whole of BEEBS to construct a single configuration, \ttt{-Ocm3}, that performed well across the benchmark suite. To verify that our method does not simply overfit the benchmark suite we use the standard 10-fold cross-validation technique to test our method on unseen programs.

\subsection{Method}
In 10-fold cross-validation, the programs are partitioned into ten training and test folds. In each fold, 90\% of the programs form the training set and the remaining 10\% form the test set. Each program appears in the test set of exactly one fold and in the training set of the other nine folds. The folds for this analysis were generated using uniform random sampling.

In each fold $x$, we construct \ttt{-Ocm3-fold-x} based on the training set and test its performance on the test set. 

\subsection{Cross-validation Results}

In cross-validation \ttt{-Ocm3} performed 4\% better than \ttt{-O3} overall and fifteen programs reached speed-ups of over 20\%~(\cref{fig:test}). In many cases, performance was close to the maximum known potential gain. Figure \ref{fig:test} also compares performance to Milepost discussed later in~\cref{sec:rel:ml}.

Three programs (\ttt{recursion}, \ttt{fac} and \ttt{ud}) ran over 20\% slower than \ttt{-O3}. This is actually an artefact of using cross-validation as each of the three programs have unique optimization requirements that are not captured by any other program in the training set. Therefore excluding these programs from the training set prevents their requirements being included in the configuration. The first two programs also feature recursive calls, which would not normally be used on embedded systems due to memory constraints.

Both \ttt{recursion} and \ttt{ud} appeared in the same cross-validation test fold. The configuration generated for this fold disables two flags (\ttt{-ftree-reassoc} and \ttt{-fipa-cp-clone}) that significantly optimize \ttt{recursion} and \ttt{ud} respectively. This is the only fold that disables these flags therefore we conclude that \ttt{recursion} and \ttt{ud} are unique in their dependence on these flags and none of the remaining training programs could prevent them from being disabled.

A similar story holds for \ttt{fac} in another fold. This program gains significant benefit from enabling \ttt{-foptimize-sibling-calls} and disabling \ttt{fmodulo-sched} but the configuration constructed in this fold disables the former and enables the latter. As in the previous scenario, this is the only fold that features these particular settings.

BEEBS was deliberately designed to include a diverse range of benchmarks with little redundancy between them, therefore we cannot expect optimal performance when training on a subset of the benchmarks. However, these results do show that our method performs well and not due purely to chance.

In conclusion, the majority of programs performed as well as or better than \ttt{-O3}. In practice, should a program perform worse than \ttt{-O3}, the user can simply choose \ttt{-O3} instead. This is a much less time-intensive task than choosing from hundreds of configurations.

\section{Testing on Other Platforms}
\label{sec:plat}

In order to demonstrate that our method can also optimize GCC for other embedded platforms we construct and test two new optimization levels \ttt{-Oavr}~(\cref{fig:oavrConfig}) and \ttt{-Oca8}~(\cref{fig:oca8config}) for the AVR and CA8 processors. We used threshold $t=5\%$ to produce these configurations but it is possible that other thresholds may improve the results further. We also ran time-intensive CE experiments on each benchmark on the two platforms to quantify the potential gains.\footnote{The CA8 analysis excludes two benchmarks and the AVR analysis excludes 23 benchmarks that do not run on these platforms.}

\begin{figure*}
\begin{center}
\includegraphics[width=\textwidth,keepaspectratio]{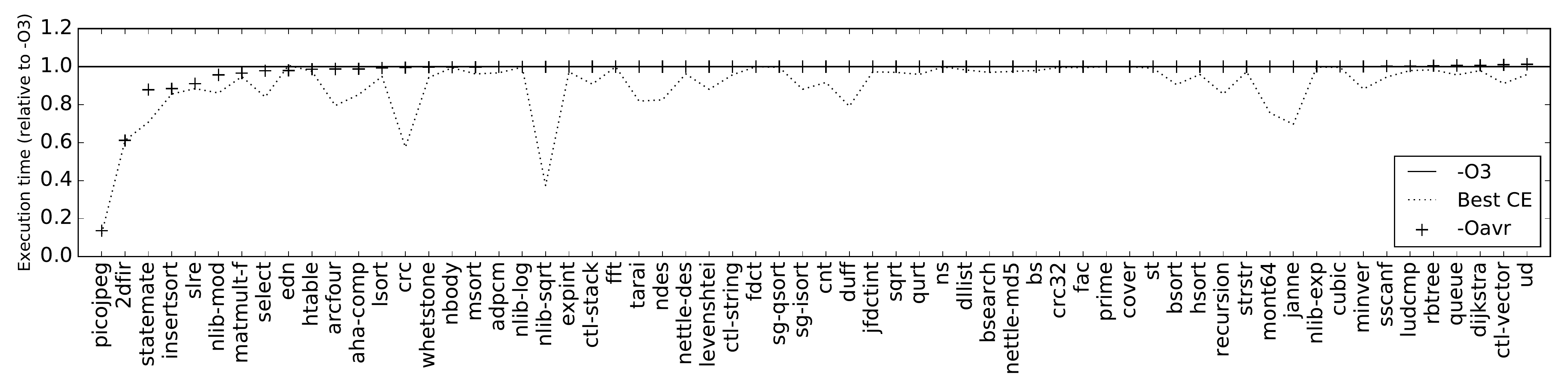}
\end{center}
\caption{Performance of \ttt{-Oavr} on the AVR}
\label{fig:avr}
\end{figure*}

The \ttt{-Oavr} configuration improves overall performance on the AVR by 3\% compared to \ttt{-O3}. Five benchmarks performed over 5\% faster than \ttt{-O3} and only two benchmarks performed very slightly worse than \ttt{-O3} (\cref{fig:avr}). The configuration disables nine flags and enables six others.

\begin{figure*}
\begin{center}
\includegraphics[width=\textwidth,keepaspectratio]{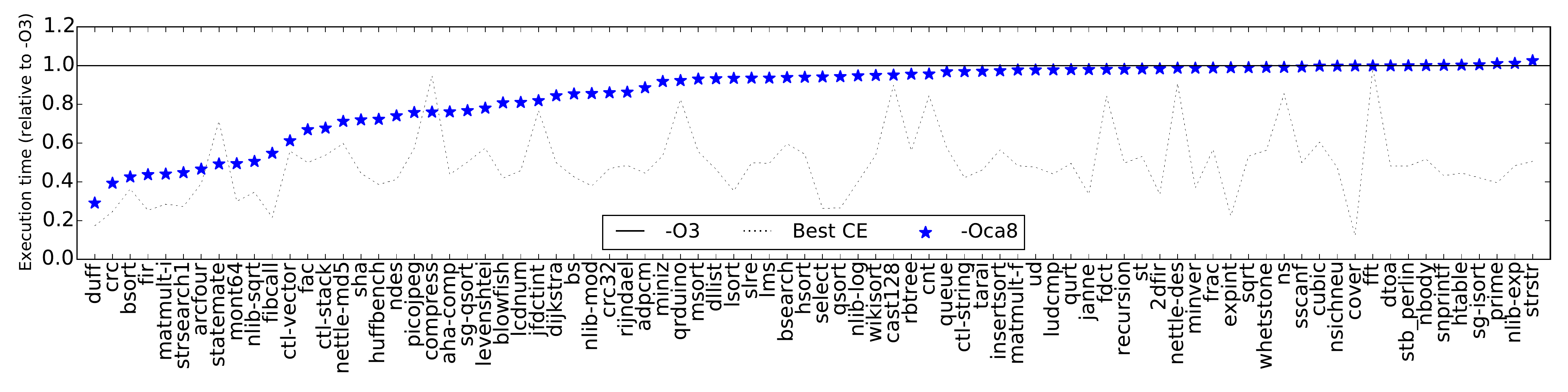}
\end{center}
\caption{Performance of \ttt{-Oca8} on the Cortex-A8}
\label{fig:ca8}
\end{figure*}

On the CA8, \ttt{-Oa8} improves overall performance by 15\% compared to \ttt{-O3}. Over half of the benchmarks performed over 5\% faster than \ttt{-O3} and only one performed slightly worse (\cref{fig:ca8}). The configuration disables 22 flags and enables five others.

As the complexity of the hardware increases we observe that the potential gains over \ttt{-O3} also increase. The AVR is a simple 8-bit processor with relatively little room for further optimization in many cases. The CA8 is the most complex of the three architectures (with its superscalar pipeline, caches and SIMD unit) and shows the most potential gains.

Several flags are common to both \ttt{-Ocm3} and \ttt{-Oca8} and therefore particular flags may have similar effects on processors from closely related families. Conversely, there is less overlap between these configurations and \ttt{-Oavr} which demonstrates that the impact of the flags is indeed dependent on the platform as well as the program.

\begin{figure}[h]
\begin{verbatim}
-O3 -fno-toplevel-reorder -fno-predictive-commoning
-fipa-pta -fgcse-sm -fno-forward-propagate
-fconserve-stack -fno-ipa-sra -ftree-loop-distribution
-fno-tree-dse -fgcse-las -fno-common
-fno-tree-scev-cprop -fno-inline-functions-called-once
-fdata-sections -fno-merge-constants
\end{verbatim}
\caption{The \ttt{-O\lowercase{avr}} Configuration}
\label{fig:oavrConfig}
\end{figure}

\begin{figure}[h]
\begin{verbatim}
-O3 -fno-tree-loop-if-convert -fno-split-wide-types
-fno-tree-cselim -fno-ipa-pure-const
-fno-tree-slp-vectorize -fno-tree-dse
-fno-tree-loop-im -fno-merge-constants
-fno-common -fconserve-stack -fno-caller-saves
-fno-tree-tail-merge -fno-inline-functions-called-once
-funroll-loops -fgcse-las -fno-cse-follow-jumps
-fno-sched-dep-count-heuristic -fno-tree-phiprop
-fno-tree-slsr -funroll-all-loops
-fno-tree-loop-distribute-patterns
-fno-tree-coalesce-vars -fno-reorder-functions
-fno-peephole2 -fno-sched-last-insn-heuristic
-fno-ipa-sra -fsched-spec-load
\end{verbatim}
\caption{The \ttt{-O\lowercase{ca}8} Configuration}
\label{fig:oca8config}
\end{figure}

\section{Related Work}
\label{sec:rel}

This section begins with a discussion on iterative compilation studies relative to our work, with particular focus on a related comparison between CE and RIC (\cref{sec:rel:iter}). Then we compare our \ttt{-Ocm3} cross-validation results to a state-of-the-art predictive approach and suggest improvements to that approach~(\cref{sec:rel:ml}).

\subsection{Iterative Compilation}
\label{sec:rel:iter}

Cavazos et al. also compared RIC and CE but in contrast to our study~(\cref{sec:invest}) they found that RIC outperformed CE. A direct comparison between the two studies is not possible as they are based on different platforms, benchmarks, optimizations and compilers. However, we give below a brief insight into the impact of flag dependencies which presents a plausible reason why CE performed better on our setup.

Combined Elimination takes $N+1$ configurations to test the performance of the initial baseline and disabling each of the $N$ flags individually. Our results show significant gains even in this initial stage of removing single flags (\cref{sec:invest:ceVSrandom}). Conversely, in \cite{cavazos2007} the majority of gains only occurred once the algorithm had begun disabling multiple flags. Therefore, flag dependencies may have a greater impact on their setup.

We suggest two ways in which the experimental setup might influence the amount of flag dependencies. Firstly, the flags of the PathScale EKOPath compiler used by~\cite{cavazos2007} may have more interdependencies than in GCC and secondly, the platform and/or benchmarks may be more sensitive to these flag dependencies.

Purini et al.~\cite{purini2013} used iterative compilation to identify a set of ten configurations such that it contains at least one good configuration for each benchmark. Once generated, these ten configurations can then be used for the iterative compilation search on new programs. The method used iterative compilation approaches to find a good configuration for each benchmark. These configurations were then pooled together and downsampled into the ten configurations. This top-down approach differs from our bottom-up approach which performs a directed search, dependent on the performance of all benchmarks, towards a single high-performing configuration.

Pallister et al~\cite{pallister2013} analyzed iterative compilation data to quantify the impact of individual flags on energy consumption of the CM3 using 82 flags from GCC 4.7 and an early version of BEEBS which contained 10 benchmarks. The study identifies the top three most significant flags for the energy consumption of each program and overall the list includes two flags from \ttt{-Ocm3}. The experiments excluded flags enabled at \ttt{-O0} and those not enabled \ttt{-O3}, therefore, many flags such as \ttt{-fcommon} and \ttt{-ftree-loop-if-convert} (which are both enabled at \ttt{-O0}) do not feature in the study.

\subsection{Machine Learning}
\label{sec:rel:ml}

We compared our \ttt{-Ocm3} cross-validation results with the state-of-the-art 1NN probabilistic machine learning approach from Milepost~\cite{fursin2011} using the same cross-validation folds. 

We produced training data for 1NN by extracting the feature vector for the most time consuming function of each program (using Milepost GCC) and combining this with the RIC data from our investigatory study~(\cref{sec:invest}). We created our own implementation of the 1NN algorithm as Milepost is not trained for the CM3 and there were difficulties in supplying our own data to the system.

Milepost 1NN performed 43\% slower than \ttt{-O3} with the majority of programs performing worse than both \ttt{-O3} and \ttt{-Ocm3} (\cref{fig:test}). Based on insights from our work, we anticipate the original Milepost 1NN approach can be improved by training with CE data (rather than RIC) and using the feature vector of the entire program (rather than the most time consuming function) to make predictions.

We tested these suggestions using 10-fold cross-validation and found that they do indeed improve the performance of 1NN on BEEBS by reducing the overall execution time by 1\% compared to \ttt{-O3}. In spite of these improvements, our new optimization level, \ttt{-Ocm3}, still performs best and our method has none of the overheads and complexities of a machine learning based approach.

Blackmore et al.~\cite{Blackmore2015} also tested 1NN on BEEBS and the CM3 and found that it performed slower than \ttt{-O3} overall. They also used Milepost's feature vector to demonstrate the wide diversity of the programs in BEEBS.

Despite several proposed machine learning approaches, there does not yet exist a direct comparison between all methods to determine the best. Such a comparison is difficult due to the lack of available and maintained implementations and training data for each approach. Furthermore, each study uses different benchmarks, platforms, compilers and optimization settings.

We have contributed to a state-of-the-art embedded benchmark suite which other studies can use to compare to our work. This paper also publishes each configuration produced by our method~(\cref{fig:ocm3config,fig:oca8config,fig:oavrConfig}) which allows future work to compare with our approach and software developers to use these configurations in practice.

\section{Conclusion and Future Work}
\label{sec:conclusion}

We have demonstrated an automatic method for tuning the GCC compiler to a given target architecture. Using our approach we generated three new optimization levels, \ttt{-Ocm3}, \ttt{-Oca8} and \ttt{-Oavr}, that outperform GCC's highest safe optimization level \ttt{-O3} on the ARM Cortex-M3, ARM Cortex-A8 and 8-bit AVR respectively.

We offer these new optimization levels as platform-specific alternatives to \ttt{-O3}. In situations where they might be found to reduce performance the user can simply opt for \ttt{-O3}. Choosing between two configurations is much less arduous than the hundreds considered by iterative compilation searches for each new program. We have shown that while our new optimization levels offer significant improvements on many benchmarks they do not guarantee the full potential gains of a time-intensive iterative compilation search tailored to a given program. Therefore, the user must decide whether it is worthwhile and feasible to invest considerable extra time in running iterative compilation to optimize their program of choice or simply use \ttt{-Ocm3}, \ttt{-Oca8} or \ttt{-Oavr}. In any case, it is feasible to try these configurations on any program developed for the CM3, CA8 or AVR.

Our approach was demonstrated on CE, but in principle, it can be applied to any iterative compilation method by changing the goal from optimizing the performance of a single program to optimizing the performance of a representative benchmark suite. We anticipate that some iterative compilation approaches may be more suited to particular benchmarks, compilers, platforms and optimizations.

In conclusion, our approach offers an automatic method to tune compilers to new architectures. Many of the gains are captured by our new method, but there is also the opportunity to run iterative compilation starting from our configuration or further enhance performance using machine learning.

In theory, compiler designers can adjust optimization levels for each architecture. Our analyses of the two flags in~\cref{sec:results} shows it is possible to reason by hand about which flags need to be removed. In practice this happens to some extent, but the fact that these flags were not removed from \ttt{-O3} for these architectures shows there is a need for automated analyses like the one we have developed.

In future work, we plan to validate our methods on more architectures and extend the evaluation to real world applications. In principle, our approach could also be applied to customize compiler settings for other compilers, metric(s) and/or specific classes of programs.

\bibliographystyle{ACM-Reference-Format}
\bibliography{ocm3} 

\end{document}